\documentclass[aps,prl,reprint,footinbib,floatfix]{revtex4-1}
\usepackage{amsmath}
\usepackage{amssymb}
\usepackage{amsthm}
\usepackage{graphicx}

\def\be{\begin{equation}}
\def\ee{\end{equation}}
\def\ber{\begin{eqnarray}}
\def\eer{\end{eqnarray}}
\def\bern{\begin{eqnarray*}}
\def\eern{\end{eqnarray*}}

\def\rv{\mathbf{r}}

\def\Gv{\mathbf{G}}
\def\kv{\mathbf{k}}

\def\qv{\mathbf{q}}

\def\0v{\mathbf{0}}
\def\1v{\mathbf{1}}
\def\2v{\mathbf{2}}
\def\3v{\mathbf{3}}

\begin{document}

\title{Optics of semiconductors and insulators: Role of local-field effects  revised}

\author {V. U. Nazarov}
\affiliation{Research Center for Applied Sciences, Academia Sinica, Taipei 11529, Taiwan}
\email{nazarov@gate.sinica.edu.tw}

\author{S. Kais}
\affiliation{Department of Chemistry,  Physics and Birck Nanotechnology Center, Purdue University, West Lafayette, IN 47907 USA}
\affiliation{Qatar Environment and Energy Research Institute, Qatar Foundation, Doha, Qatar}

\begin{abstract}
We show that by nullifying  the short-wave response to the long-wave excitation (local-field-effects),
the adiabatic time-dependent density-functional theory (TDDFT) of optics of semiconductors and insulators
can be brought into excellent agreement with experiment. This indicates
that the wing elements [($\Gv,\0v)$ and $(\0v,\Gv)$, $\Gv\ne\0v$] 
of both the Kohn-Sham (KS) density-response function $\chi^s$ and 
the exchange-correlation kernel $f^{xc}$ are greatly overestimated by the existing approximations to  the 
static DFT and TDDFT, respectively,
to the extent that zero is a better approximation for them
than the corresponding values provided by current theories.
The head element of $f^{xc}$ is thereby fixed by the static macroscopic dielectric constant $\epsilon_M$.
Our method yields accurate optical spectra including both the weakly and strongly bound excitons,  while
its computational cost is extremely low, since only the head element of the KS response
matrix and the static dielectric constant are needed.
\end{abstract}


\maketitle

It is known since works of Adler \cite{Adler-62} and Wiser \cite{Wiser-63} that in order
to obtain the macroscopic dielectric function  $\epsilon_M(\qv,\omega)$ of a crystal, 
one must invert the microscopic dielectric matrix
$\epsilon_{\Gv\Gv'}(\qv,\omega)$ indexed with the reciprocal lattice vectors.
Then 
\begin{equation}
\epsilon_M(\qv,\omega)=\frac{1}{\epsilon^{-1}_{\0v\0v}(\qv,\omega)}.
\label{AW}
\end{equation}
As a result, in general, $\epsilon_M(\qv,\omega)\ne\epsilon_{\0v\0v}(\qv,\omega)$,
which fact is due to the short-wave response to the long-wave perturbation and is usually
referred to as the local-field effects (l.f.e.) (see, e.g., Ref. \onlinecite{Sturm-82} 
and references therein).

The time-dependent density-functional theory (TDDFT) \cite{Gross-85}, which has become a preferential 
approach
in the studies of dynamic quantum-mechanical processes in general, and 
in optics, in particular \cite{Kim-02,Kim-02-2,Sottile-03,Botti-04,Sharma-11,Nazarov-11,Yang-12,Bates-12,Yang-13,Trevisanutto-13}, 
takes full care of  l.f.e., representing crystals' response functions with matrices indexed with reciprocal lattice vectors. 
The key quantity of TDDFT is the exchange-correlation kernel $f^{xc}$, which, together with the Kohn-Sham (KS)
single-particle density-response function $\chi^s$, determine the interacting-particles
density-response function $\chi$ through the equality \cite{Gross-85}
\begin{equation}
\chi^{-1}_{\Gv\Gv'}(\qv,\omega)=(\chi^s)^{-1}_{\Gv\Gv'}(\qv,\omega)
-\frac{4\pi}{|\Gv+\qv|^2} \delta_{\Gv\Gv'}- f^{xc}_{\Gv\Gv'}(\qv,\omega).
\label{8}
\end{equation}
While $\chi^s$ is constructed using the single-particle states obtained
with a given approximation to the static exchange-correlation potential $v_{xc}(\rv)$ \cite{Kohn-65},
$f^{xc}$ is a true many-body quantity containing, in principle exactly, all the dynamic exchange-correlation effects
in a real interacting system.

A great amount of efforts has been invested into the development of approximations to  $f^{xc}$
of crystalline semiconductors and insulators
\cite{Kim-02,Kim-02-2,Sottile-03,Botti-04,Kim-02-2,Sharma-11,Nazarov-11,Yang-12,Bates-12,Yang-13,Trevisanutto-13}.
They range from the computationlly demanding ones which provide little gain in the efficiency compared
with the solution of Bethe-Salpeter equation \cite{Albrecht-98}, although solidly grounded 
theoretically \cite{Kim-02,Sottile-03},
to very practicable {\it ad hoc} schemes \cite{Sharma-11}.
In this Letter, 
which tends to the latter category,
we come up with a simple {\it ansatz} which leads to an  approximation by far simpler and  computationally more
efficient than any of the existing approaches. At the same time our method provides very accurate
optical spectra of semiconductors and insulators including, in particular, the weakly and strongly bound excitons.
Specifically, we nullify  the l.f.e., in other words, the contribution from the wing elements of both $\chi^s$ and $f^{xc}$ are set to zero
\begin{align}
&\lim_{\qv\rightarrow 0} \frac{\chi^s_{\Gv\ne\0v,\0v}(\qv,\omega)}{G q} = 0,
\label{chisnull}\\
&\lim_{\qv\rightarrow 0} G q f^{xc}_{\Gv\ne\0v,\0v}(\qv,\omega) = 0.
\label{fxcnull}
\end{align}
From the expression for the microscopic dielectric matrix
\begin{equation}
\epsilon^{-1}_{\Gv\Gv'}(\qv,\omega)=\delta_{\Gv\Gv'}+\frac{4\pi}{|\Gv+\qv| |\Gv+\qv'|} \chi_{\Gv\Gv'}(\qv,\omega),
\label{epsMM}
\end{equation}
equation~(\ref{8}), and the mathematical fact that the inverse of a matrix with zero wings
is a matrix with zero wings, we see that Eqs.~(\ref{chisnull})
and (\ref{fxcnull}) lead to
\begin{equation}
\epsilon_{\Gv\ne\0v,\0v}(\qv= \0v,\omega)=0.
\label{epsMM0w}
\end{equation}
Obviously, any two of the Eqs.~(\ref{chisnull}),
(\ref{fxcnull}), and (\ref{epsMM0w}) entail the third one.
Equations (\ref{AW}) - (\ref{fxcnull})  also yield
\begin{equation}
\lim_{\qv\rightarrow \0v} q^2 f^{xc}_{\0v\0v}(\qv,0)=
\lim_{\qv\rightarrow \0v}
\frac{q^2} {\chi^s_{\0v\0v}(\qv,0)}-\frac{4\pi}{1-\epsilon_M},
\label{fxchead2}
\end{equation}
where $\epsilon_M$ is the macroscopic static dielectric function
\footnote{In this context, 'static' means the so called high-frequency dielectric constant
$\epsilon_M\equiv \epsilon_\infty=\lim_{\omega\rightarrow 0} \epsilon_M(\qv=\0v,\omega)$.
}.
Taking use of the {\em adiabatic} TDDFT,
we extend Eq.~(\ref{fxchead2}) to finite frequencies
\begin{equation}
\lim_{\qv\rightarrow \0v} q^2 f^{xc}_{\0v\0v}(\qv,\omega)=
\lim_{\qv\rightarrow \0v}
\frac{q^2} {\chi^s_{\0v\0v}(\qv,0)}-\frac{4\pi}{1-\epsilon_M}.
\label{fxchead2w}
\end{equation}
Since all the matrices involved have zero wings, no body elements are now relevant to the calculation
of the macroscopic dielectric function $\epsilon_M(\omega)$,
and Eqs.~(\ref{chisnull}), (\ref{fxcnull}), and (\ref{fxchead2w})
constitute a closed-form solution as soon as the static $\epsilon_M$ is known.
The latter can be found by an independent calculation or taken from experiment.

In Fig.~\ref{fig_chch0}, we present results  for the optical absorption 
of several semiconductors and insulators obtained with the use of Eqs.~(\ref{chisnull}),
(\ref{fxcnull}), and (\ref{fxchead2w}).
Calculations were carried out with the full-potential linear augmented plane-waves (LAPW)  code Elk \cite{Elk}. 
We use Tran and Blaha's  meta generalized gradient approximation (meta-GGA)  (TB09) 
for the exchange potential, which provides realistic band-gaps \cite{Tran-09}. For correlations, 
the local-density approximation (LDA) potential \cite{Perdew-92} is used.
Convergence was achieved with the shifted $32\times 32\times 32$ $\kv$-points grid and the reciprocal
vector cut-off $G=12$ bohr$^{-1}$. 
Clearly, the overall agreement between the theory and experiment
is very good: The positions of the excitonic features in the spectra are correct 
for all the considered materials and their intensity compared with other peaks 
is mostly accurate too. Figure~\ref{fig_chch0r} shows the real part of the 
dielectric function of the same crystals.

It must be noted that the idea to relate the head element of $f^{xc}$ to the static macroscopic dielectric function
dates back to Ref.~\onlinecite{Botti-04}. 
Our results, however, show that this idea is quantitatively successful only if the l.f.e. are discarded,
otherwise there is no good agreement between theory and experiment \cite{Botti-04}.
Besides, only if there are no wing elements Eq.~(\ref{fxchead2}) holds,
providing a basis for the adiabatic TDDFT within this approach.

In conclusion, 
we suggest the simplest of the existing approximations
in the time-dependent density-functional theory of optics of semiconductor and insulators,
i.e., the approximation of the zero wing elements of  the response matrices.
This  proves to be remarkably successful in reproducing
the experimental optical spectra, including both weakly and strongly bound excitons.  
The simplicity of the implementation combined with the high accuracy 
has the potential of making this method a useful theoretical tool
in optics and possibly beyond.

\acknowledgments
V.U.N. acknowledges partial support from National Science Council, Taiwan, Grant No. 100-2112-M-001-025-MY3
and he is grateful for the hospitality of Qatar Energy and Environment Institute, Qatar Foundation, Qatar.


%

\begin{widetext}

\begin{figure}[h] 
\includegraphics[width=1.0 \columnwidth,clip=true,trim=10 0 0 0]{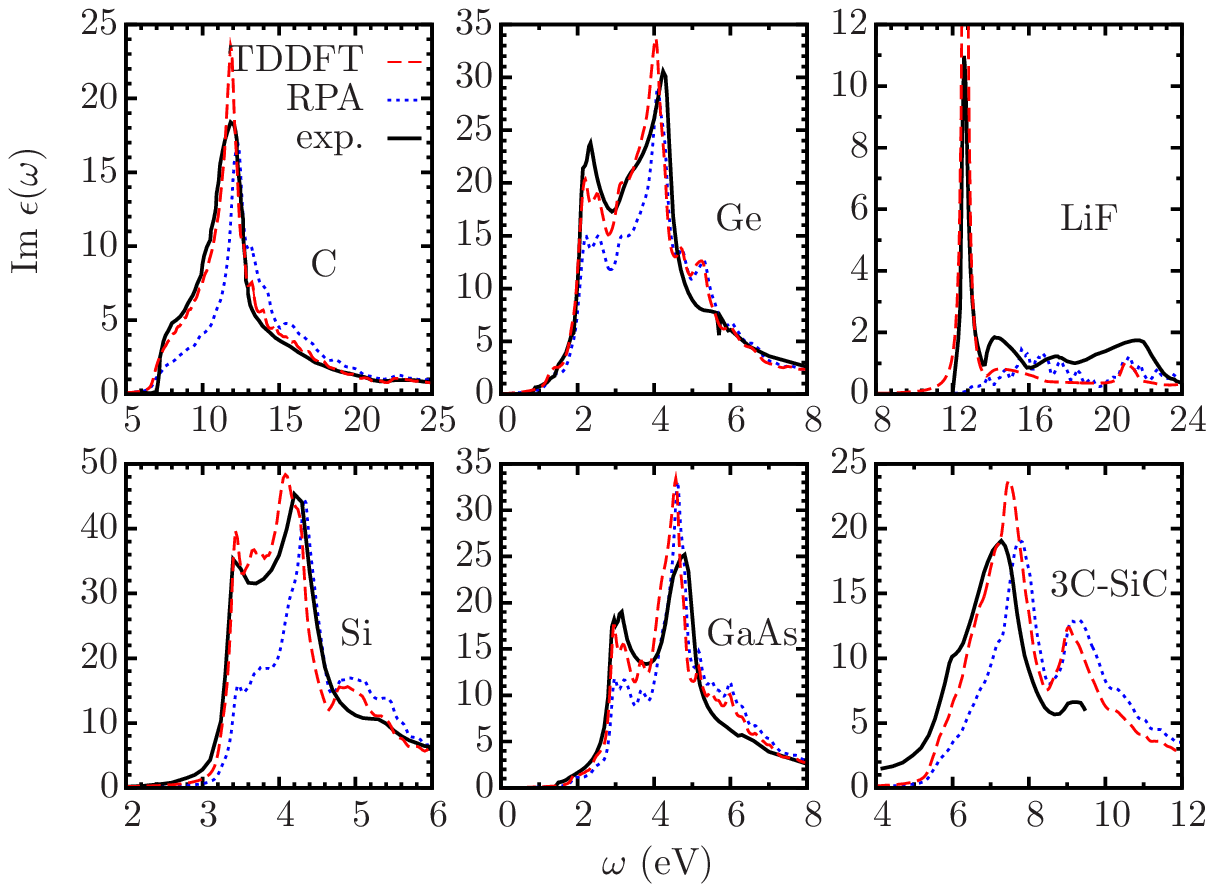} 
\caption{\label{fig_chch0} (color online) 
Imaginary part of the dielectric function of several semiconductors and insulators.
The theory (dashed red line) which implements Eqs.~(\ref{chisnull}), (\ref{fxcnull}), and (\ref{fxchead2w}) is used.
The single-particle
band structure and, consequently, $\chi^s_{\Gv\Gv'}(\0v,\omega)$
were calculated with MGGA exchange 
and LDA correlation potentials.
Blue dotted line is random phase approximation (RPA) with meta-GGA ground-state calculation.
Experiment (black solid line) is from Ref.~\onlinecite{Palik}
for all species except for  SiC, which is of Ref.~\onlinecite{Logothetidis-96}.
Static dielectric constants $\epsilon_M$ used in Eq.~(\ref{fxchead2w})
are 5.6, 11.7, 16, 10.9, 1.92, and 6.52 for diamond, Si, Ge, GaAs, LiF, and SiC, respectively.
}
\end{figure}

\begin{figure}[h] 
\includegraphics[width=1.0 \columnwidth,clip=true,trim=10 0 0 0]{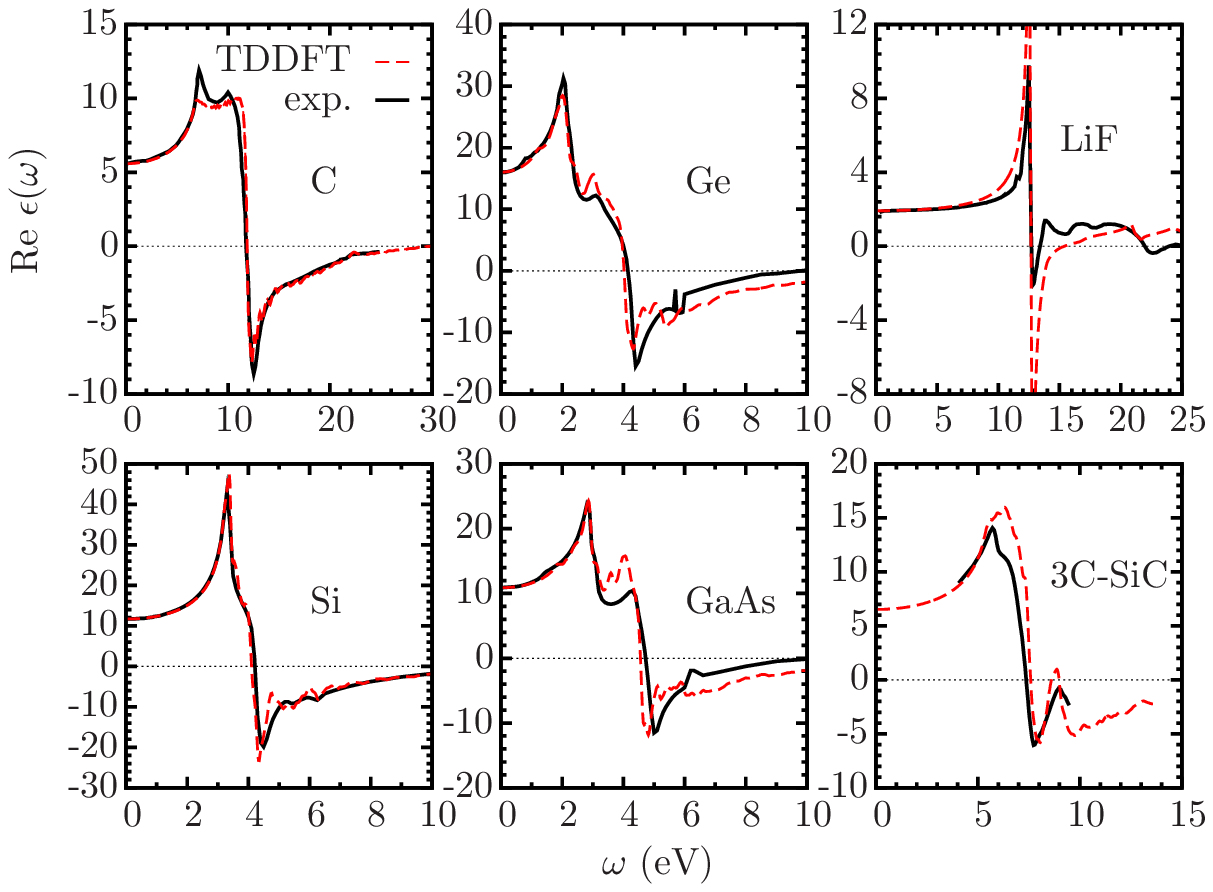} 
\caption{\label{fig_chch0r} (color online) 
Real part of the dielectric functions of the same crystals as in Fig.~\ref{fig_chch0}.
}
\end{figure}

\end{widetext}

\end{document}